# Coordinated Charging and Discharging Strategies for Plug-in Electric Bus Fast Charging Station with Energy Storage System

Huimiao Chen, Zechun Hu [*], Hongcai Zhang, Haocheng Luo

Department of Electrical Engineering, Tsinghua University, Beijing, 100084, China
[*]zechhu@tsinghua.edu.cn

**Abstract:** Plug-in electric bus (PEB) is an environmentally friendly mode of public transportation and plug-in electric bus fast charging stations (PEBFCSs) play an essential role in the operation of PEBs. Under effective control, deploying an energy storage system (ESS) within a PEBFCS can reduce the peak charging loads and the electricity purchase costs. To deal with the (integrated) scheduling problem of (PEBs charging and) ESS charging and discharging, in this study, we propose an optimal real-time coordinated charging and discharging strategy for a PEBFCS with ESS to achieve maximum economic benefits. According to whether the PEB charging loads are controllable, the corresponding mathematical models are respectively established under two scenarios, i.e., coordinated PEB charging scenario and uncoordinated PEB charging scenario. The price and lifespan of ESS, the capacity charge of PEBFCS and the electricity price arbitrage are considered in the models. Further, under the coordinated PEB charging scenario, a heuristics-based method is developed to get the approximately optimal strategy with computation efficiency dramatically enhanced. Finally, we validate the effectiveness of the proposed strategies, interpret the effect of ESS prices on the usage of ESS, and provide the sensitivity analysis of ESS capacity through the case studies.

## 1. Nomenclature

*Indices and Sets*

| | |
|---|---|
| $n$ | Index of PEB. |
| $m$ | Index of fast charging port. |
| $k$ | Index of time interval in the following optimising time horizon. |
| $i_n, j_n$ | Index of parking of PEB $n$ in the following optimising time horizon. |
| c / d | Subscript of charging/discharging. |
| $N$ | Index set of PEBs. |
| $M$ | Index set of fast charging ports. |
| $K$ | Index set of time intervals in the following optimising time horizon. |
| $P(i_n)$ | Index set of time intervals during parking $i_n$ of PEB $n$. |
| $I(n)$ | Index set of parking of PEB $n$ in the following optimising time horizon. |
| $P(I(n))$ | Index set of time intervals in the following optimising time horizon when PEB $n$ is expected to be parking. |
| $\text{card}(X)$ | The number of elements in set $X$. |

*Parameters and Variables*

| | |
|---|---|
| $S_n^{\text{PEB}}$ | Battery capacity of PEB $n$ (kWh). |
| $S^{\text{ESS}}$ | Energy capacity of ESS (kWh). |
| $P_c^{\text{PEB}}$ | Rated charging power of PEBs (kW). |
| $P_{\text{peak}}$ | Peak load of PEBFCS (kW). |
| $P_{c,\max}^{\text{ESS}}$ | Maximum charging power of ESS (kW). |
| $P_{d,\max}^{\text{ESS}}$ | Maximum discharging power of ESS (kW). |
| $\eta_c^{\text{PEB}}$ | Charging efficiency of PEBs. |
| $\eta_c^{\text{ESS}}$ | Charging efficiency of ESS. |
| $\eta_d^{\text{ESS}}$ | Discharging efficiency of ESS. |
| $SOC_{\min}^{\text{PEB}}$ | Minimum state of charger (SOC) for PEB batteries. |
| $SOC_{\min}^{\text{ESS}}$ | Minimum SOC for ESS. |
| $SOC_{n,i_n}^{\text{PEB}}$ | SOC of $i_n$th arrival of PEB $n$. |
| $SOC_k^{\text{ESS}}$ | SOC of ESS at the beginning of time interval $k$. |
| $\Delta SOC_{n,i_n}^{\text{PEB}}$ | SOC difference of PEB $n$ between $i_n$th departure and the next return. |
| $a_{n,i_n}$ | Time interval of $i_n$th expected return of PEB $n$ ($a_{n,i_n} \in K$). |
| $l_{n,i_n}$ | Time interval of $i_n$th expected departure of PEB $n$ ($l_{n,i_n} \in K$). |
| $L_k$ | Power of other loads excluding PEB charging loads in time interval $k$ (kW). |
| $\Delta t$ | Duration of a time interval (min). |
| $\pi_k^{\text{TOU}}$ | Electricity price in time interval $k$ (RMB/kWh). |
| $\pi^{\text{ESS}}$ | Price of ESS (RMB/kWh). |
| $\pi^{\text{Cap}}$ | Capacity charge of PEBFCS (RMB/kW). |
| $n^{\text{ESS}}$ | The number of charge-discharge cycle of ESS. |
| $\alpha$ | Discount rate of the capacity charge (%). |
| $\gamma$ | Life cycle of PEBFCS (year). |
| $\mathbf{C}^{\text{PEB}}$ | Charging state matrix of PEBs (dimensions: $\text{card}(N) \times \text{card}(K)$). |
| $\mathbf{C}^{\text{FCP}}$ | Charging state matrix of fast charging ports (dimensions: $\text{card}(M) \times \text{card}(K)$). |
| $\mathbf{P}_c^{\text{ESS}}$ | Vector of charging power of ESS (dimensions: $1 \times \text{card}(K)$). |
| $\mathbf{P}_d^{\text{ESS}}$ | Vector of discharging power of ESS (dimensions: $1 \times \text{card}(K)$). |
| $c_{n,k}^{\text{PEB}}$ | Element of $\mathbf{C}^{\text{PEB}}$, binary variable, 1: on |
| $c_{m,k}^{FCP}$ | charge; 0: off charge.<br>Element of $\mathbf{C}^{FCP}$, binary variable, 1: on state; 0: off state. |
| $P_{c,k}^{EES}$ | Element of $\mathbf{P}_c^{ESS}$, the charging power of ESS in time interval $k$ (kW). |
| $P_{d,k}^{EES}$ | Element of $\mathbf{P}_d^{ESS}$, the discharging power of ESS in time interval $k$ (kW). |
| $u_{n,k}, v_{n,k}$ | Auxiliary variables. |

## 2. Introduction

The wide use of fossil energy has resulted in global warming and severe environmental pollution [1]. Plug-in electric vehicles (PEVs) have incomparable advantage over fuel-powered vehicles in environmental protection and sustainable development [2], [3]. With development and popularization of PEVs, a large-scale of PEVs will be connected to the public power grid in the future. The incremental charging loads of PEVs will have a massive impact on the existing power system [4], [5]. For example, difference between load peak and off-peak may increase; power quality could be deteriorated; distribution networks will face new challenges, including increasing of network losses [6], [7], overloading of transformers [8], [9], excessively heavy line loads and larger voltage deviations [10], [11], etc.

Present research has shown coordinated charging of PEVs is able to effectively reduce the negative impact of PEVs' charging loads on the power system [12]-[17]. The optimisation objectives and methods of PEV coordinated charging are various in the literature. Under the time-of-use (TOU) prices, reference [12] proposes a cost-optimal control strategy for multiple PEV aggregators to guarantee that the distribution system runs within the security limits. In deregulated electricity market, authors of [13] present an optimal charging control method for PEVs to provide ancillary services based on the forecast of future electricity prices. In [14], a two-stage optimisation method is developed: Firstly, in order to achieve peak shaving and valley filling, PEVs are allocated appropriate charging periods according to the urgency degree; then the charging sites are optimised to minimise transmission losses. In [15], a threshold admission and greedy scheduling policy is proposed to maximise the revenue of charging services for large-scale electric vehicles. Taking vehicle to grid into account, reference [16] builds an optimal scheduling model to minimise the energy consumption and carbon emission, whereas reference [17] formulates a dynamic charging control strategy for providing frequency regulation services.

Though a lot of research on PEV coordinated charging has been done, most of the relevant works concentrate on studies of slow or normal charging mode of private PEVs, which are not applicable for a plug-in electric bus fast charging station (PEBFCS). As a special type of PEVs, plug-in electric bus (PEB) is an electric bus which is powered by electricity and can be recharged from an external source of electricity. In [18], an effective charging strategy for PEBFCSs is proposed to minimise the power purchase costs by responding to the TOU prices, and as the result, the peak loads are mitigated as well. But the work is not suitable for a PEBFCS with ESS. Nowadays, with the rapid development of energy storage technology, installing ESS in the charging station can achieve better demand response [19]. However, only a few published literature focuses on charging stations with ESS. Reference [20] proposes a control strategy for PEV fast charging station equipped with a flywheel ESS, which is able to work without any digital communication between the grid-tied and flywheel ESS converters. Reference [21] provides a method to schedule PEV charging with energy storage and shows that aggregator's revenue varies as the number of PEVs and the number of energy storage units change. Authors of [22] present a coordinated control strategy for ESS to reduce the electricity purchase costs and flatten the charging load profile. However, the investment costs of energy storage are not taken into account both in [21] and [22]. Besides, the original load curve is given and fixed in [22] so that the elasticity of PEV charging loads cannot play a role. In [19], the value of ESS in a PEBFCS is discussed as the core problem and the control strategy of PEBs is not concerned.

To the best of our knowledge, there is no existing papers which study the optimal load scheduling method for a PEBFCS with ESS. Thus, herein, we aim to develop an integrated control strategy for both ESS and PEB loads in a PEBFCS with ESS in order to achieve maximum economic benefits. It is worthy to note that a PEBFCS with ESS is a valuable research object for following reasons: 1) PEB is a green public means of transportation, which is convenient for centralized control and management. Moreover, a large number of PEB lines have been in commercial operation or demonstrational operation in some cities (e.g. Shenzhen and Chongqing, China). 2) Fast charging stations are regarded as the promising providers of public PEVs' charging service in the future because they can provide large charging power and meet urgent charging demands. 3) As the technology of ESS advances, the efficiency and lifespan of ESS are expected to be improved and its price declines. Thus ESS is an effective supplement for a PEBFCS to reduce the high capacity charge for the grid integration as well as to reduce the charging costs through arbitraging the price differences under TOU price scheme.

Based on the above considerations and motivations, the main procedures and contributions of the paper are summarized below.:

1) A coordinated charging strategy for PEBs without considering ESS is formulated as the baseline strategy. Additionally, under the coordinated PEB charging scenario (PEB charging loads are controllable), an optimal coordinated charging and discharging strategy involving PEBs and ESS is proposed. The control of ESS and PEBs is optimised in an integrated way and the combined control strategy achieves the best optimality.

2) Under the uncoordinated PEB charging scenario (PEB charging loads are uncontrollable), an optimal coordinated charging and discharging strategy of ESS is presented.

3) To enhance the computation efficiency, under the coordinated PEB charging scenario, a heuristics-based method is further developed to get the approximately optimal control strategy of ESS and PEBs.

4) Operation costs, load profiles and some other important indices of a given PEBFCS with ESS are simulated and compared with the ones without ESS under both two scenarios, i.e., coordinated PEB charging scenario and uncoordinated PEB charging scenario, to verify the effectiveness of the proposed control strategies. The impacts of ESS capacity on economic benefits are also analyzed.

The remainder of the paper is organized as follows. Section 3 describes the scenario for the proposed strategies. The details and the mathematical formulations of the strategies are presented in Section 4. Section 5 shows case studies and complements our analysis. Finally, Section 6 concludes.

### 3. Scenario Descriptions

In order to ensure the security of power grid, the power capacity for a PEBFCS is usually adequate for simultaneous charging of all the fast charging piles, otherwise the total power of chargers might exceed the capacity of distribution transformer and the line overload might occur. However, in the actual operation, the total charging power of the PEBFCS seldom hits the upper limit [19]. For this reason, installing ESS rather than a distribution transformer with overlarge capacity could be a more economical way by reducing the grid connection fee, i.e., the capacity charge for a PEBFCS.

The configuration of a PEBFCS with ESS is illustrated in Fig. 1. The network supplies power to the station through the local distribution transformer. And ESS, PEB charging piles and the appliances of nearby residential or commercial areas (other loads) are connected to the secondary side of the distribution transformer. If the station has exclusive distribution transformer, the ratio of the power of other loads will be approximately zero. In this text, it is assumed that PEBFCS purchases electricity from the utility at TOU electricity prices and provides fast charging service to PEBs. Note that in deregulated electricity markets, the forecasting electricity prices can take place of the TOU electricity prices and the control strategy proposed in the following still works.

For a PEBFCS, we suppose that the number of PEBs, i.e., $card(M)$, the number of fast charging piles, i.e., $card(N)$, and the PEB departure time-table are given. The control system of PEBFCS is able to acquire SOCs of PEBs when they arrive at the station through built-in battery management systems on PEBs. And $\Delta SOC_{n,i_n}^{PEB}$ can be used to forecast the following arrival SOCs. In actual operation, $\Delta SOC_{n,i_n}^{PEB}$ can be obtained based on historical data. Note that the proposed control strategy in this paper is to some degree resistant to the influence of the SOC forecasting errors because the real-time control is updated periodically and the negative effect caused by the forecasting errors will be gradually mitigated (see Section 4 for the details of the proposed control strategy). In order to reduce the negative effect on battery lifespan caused by excessive discharging, the minimum SOCs for PEB batteries and ESS are set, i.e., $SOC_{min}^{PEB}$ and $SOC_{min}^{ESS}$. Also, due to the harmful effect of fast charging start-stop on chargers and batteries, continuous (uninterruptible) charging is adopted.

For the control strategies, the optimising time horizon is discretely divided into $card(K)$ time intervals equally and the duration of each time interval is $\Delta t$. Charging power of PEB, and charging and discharging powers of ESS are regarded as constants in each time interval. Intuitively, larger $card(K)$ (smaller $\Delta t$) makes the results more accurate, but the computation burden will be heavier. In practice, the value of $card(K)$ and $\Delta t$ can be adjusted according to the accuracy requirements and the computational performances.

Based on the above scenario descriptions, the coordinated charging and discharging strategies for PEBFCS with ESS is computed through optimisation models to meet multiple constraints, such as charging demands of PEBs and continuous fast charging of PEB batteries, and to improve the economic benefit of PEBFCS, detailed in the next section.

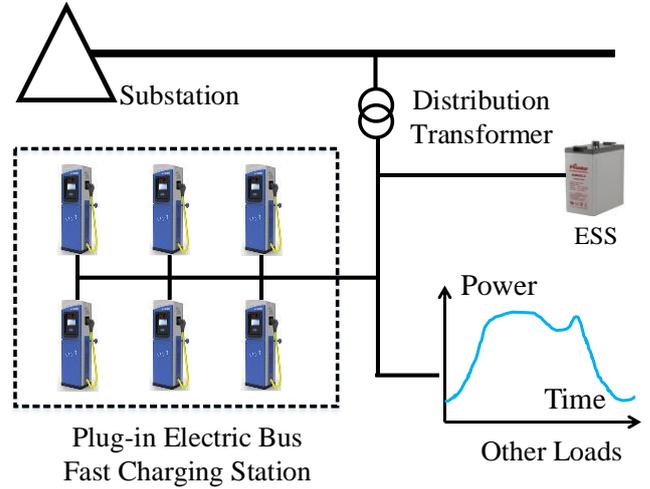

***Fig. 1.*** *Schematic illustration of a PEBFCS with ESS*

### 4. Strategies and Mathematical Formulations

#### 4.1. Control Strategy Overview

In this subsection, we outline the proposed control strategies. A rolling horizon optimisation method is applied to implementation of the proposed control strategy. Every time, a real-time control strategy is formulated according to information of the following optimising time horizon (from the next time interval to the $card(K)$ th time interval), and to ensure the real-time control performance, each control strategy will be only executed for one time interval and then a recalculated strategy will substitute for it for the next time interval.

For the control strategy under coordinated PEB charging scenario, it is operated through a three-step serial processing procedure at each time interval. When a new time interval begins, the control system will orderly 1) implement the strategy generated in the last time interval by controlling the on-off states of fast charging piles and the charging or discharging power of ESS; 2) make sure all the necessary data ready and then calculate the new charging and discharging strategy of PEBs and ESS (detailed subsequently), which will be implemented in the next time interval; 3) prepare data for the next cycle. Here, data preparation in a time interval include that: 1) if a PEB arrives, the control system will acquire the number of the PEB and read its current SOC, i.e., its $SOC_{n,1}^{PEB}$; 2) the next arrival SOCs of the rest PEBs, i.e., the other $SOC_{n,1}^{PEB}$, are estimated through the last departure SOCs (if there is communication

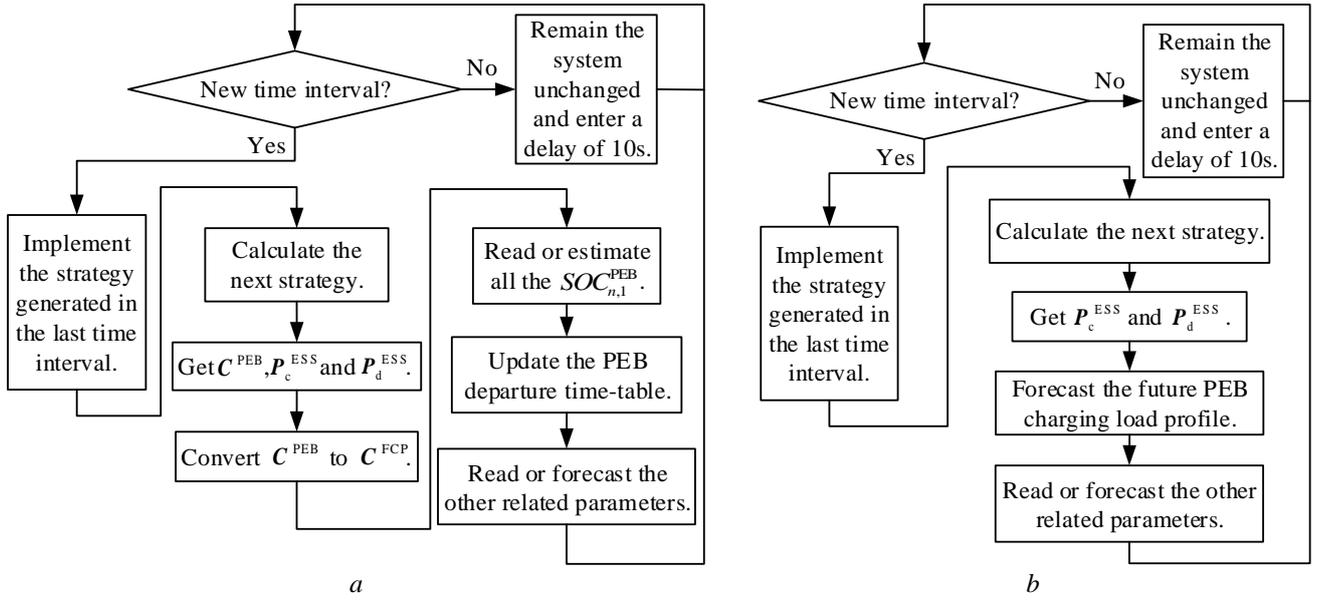

***Fig. 2.*** *Flow charts of the control strategies*
*(a)* Under coordinated PEB charging scenario, *(b)* Under uncoordinated PEB charging scenario between PEBFCS and PEBs on road, the current SOCs of all the PEBs will be available) and $\Delta SOC_{n,i_n}^{\text{PEB}}$; 3) the PEB departure time-table is updated according to the current information of PEBs; 4) the control system gets the other necessary information, which can be extracted from the database (e.g. the battery capacity $S_n^{\text{PEB}}$, the electricity price $\pi_k^{\text{TOU}}$) or be predicted based on historical data (e.g. the power of other loads $r_k$). The corresponding flow chart is shown in Fig. 2 (a).

As for the control strategy under uncoordinated PEB charging scenario, the processing procedure is similar but becomes less complicated without regard to the control of PEB charging loads. At the beginning of a new time interval, the control system will orderly 1) implement the strategy generated in the last time interval by controlling the charging or discharging power of ESS; 2) calculate the new charging and discharging strategy of ESS (detailed subsequently); 3) as the basis for the strategy formulation in the next cycle, prepare the necessary data (e.g. the electricity price $\pi_k^{\text{TOU}}$, the power of other loads $r_k$) and forecast the PEB charging load profile in the following optimising time horizon. The corresponding flow chart is shown in Fig. 2 (b).

### 4.2. Formulations of Optimisation Models

Mathematical models in this section aim at optimising the economic benefits of PEBFCS, i.e., minimising the equivalent operation costs of the following optimising time horizon.

*4.2.1 Optimisation Model for Coordinated PEB Charging Scenario without ESS:* Here, the optimisation model (*Model A*) is utilized to compute the coordinated PEB charging strategy, which is used as the reference. The objective function is to minimise the electricity purchase costs ($EPC$) and the equivalent capacity charge ($ECC$) in the following optimising time horizon. The whole model is formulated as (1)-(10).

$$\min EPC + ECC \quad (1)$$

where

$$EPC = \sum_{k \in K} \left( \sum_{n \in N} c_{n,k}^{\text{PEB}} P_c^{\text{PEB}} \right) \Delta t \pi_k^{\text{TOU}} \quad (2)$$

$$ECC = \Phi \Gamma P_{\text{peak}} \pi^{\text{Cap}},$$

$$\Phi = \frac{\text{card}(K)\Delta t}{365 \times 24 \times 60}, \Gamma = \frac{\alpha(1+\alpha)^\gamma}{(1+\alpha)^\gamma - 1} \quad (3)$$

subject to:

$$P_{\text{peak}} \geq \sum_{n \in N} c_{n,k}^{\text{PEB}} P_c^{\text{PEB}} + L_k, k \in K \quad (4)$$

$$\sum_{n \in N} c_{n,k}^{\text{PEB}} \leq \text{card}(M), k \in K \quad (5)$$

$$c_{n,k}^{\text{PEB}} = 0, k \in K \setminus P(I(n)), n \in N \quad (6)$$

$$u_{n,k} \geq c_{n,k}^{\text{PEB}} - c_{n,k-1}^{\text{PEB}}, n \in N, k \in K \setminus \{1\} \quad (7)$$

$$v_{n,k} \geq c_{n,k}^{\text{PEB}} - c_{n,k+1}^{\text{PEB}}, n \in N,$$
$$k \in K \setminus \{\text{card}(K)\} \quad (8)$$

$$u_{n,k}, v_{n,k} \in \{0,1\} \quad (9)$$

$$\sum_{k \in P(i_n)} u_{n,k} = \sum_{k \in P(i_n)} v_{n,k} \leq 1, i_n \in I(n), n \in N \quad (10)$$

$$\left( \sum_{i_n \leq j_n} \Delta SOC_{n,i_n}^{\text{PEB}} - SOC_{n,1}^{\text{PEB}} + SOC_{\min}^{\text{PEB}} \right) S_n^{\text{PEB}}$$
$$\leq \sum_{i_n \leq j_n} \sum_{k \in P(i_n)} c_{n,k}^{\text{PEB}} P_c^{\text{PEB}} \eta_c^{\text{PEB}} \Delta t \leq$$
$$\left( \sum_{i_n \leq j_n - 1} \Delta SOC_{n,i_n}^{\text{PEB}} + 1 - SOC_{n,i_n}^{\text{PEB}} \right) S_n^{\text{PEB}},$$
$$j_n \in I(n), n \in N. \quad (11)$$

In the above model, equations (2) calculates the electricity purchase costs by summing the costs in all the time intervals. In (2), $c_{n,k}^{\text{PEB}}$ denotes the charging state of PEB $n$ at time interval $k$; $P_c^{\text{PEB}}$ denotes the rated charging power of PEBs; $\Delta t$ denotes the duration of a time interval; $\pi_k^{\text{TOU}}$ is the electricity price at time interval $k$. Equation (3) calculates the equivalent capacity fees of PEBFCS. In (3), $\Gamma$ is capital recovery factor ($\alpha$ is the discount rate of the capacity charge and $\gamma$ is the life cycle of PEBFCS), which converts the initial investment costs into a stream of equal annual payments over $\gamma$ years, and the equivalent annual capacity charge is multiplied by a ratio $\Phi$, which is the proportion of the duration of optimisation time horizon to a year (see the second equation in (3)), to proportionally count the capacity charge for the following optimising time horizon. $\pi^{\text{Cap}}$ is the capacity charge of PEBFCS and the peak load $P_{\text{peak}}$ can calculated by (12), where $L_k$ is the total power of other loads excluding PEV charging loads in time interval $k$. In the model, constraints (4) are linearized expressions to describe $P_{\text{peak}}$ instead of (12) without any sacrifice of optimality, because the optimal solution must meet equation (12) to achieve the lowest costs.

$$P_{\text{peak}} = \max_{k \in K}\left(\sum_{n \in N} c_{n,k}^{\text{PEB}} P_c^{\text{PEB}} + L_k\right) \quad (12)$$

For constraints (5)-(11), (5) are the upper charging pile number constraints; constraints (6) describe that only parking PEBs can be on charge, where $P(I(n))$ is the index set of time intervals in the following optimising time horizon when PEB $n$ is expected to be parking and $P(I(n)) = \sum_{i_n \in I(n)} P(i_n) = \{a_{n,1}, \cdots, l_{n,1}\} \cup \cdots \cup \{a_{n,\text{card}(I(n))}, \cdots, l_{n,\text{card}(I(n))}\}$, $n \in N$; constraints (7)-(10) ensure the continues charging of PEBs by introducing two auxiliary variables, i.e., $u_{n,k}$ and $v_{n,k}$. Note that $u_{n,k}$ and $v_{n,k}$ guarantee the continuity of PEB charging by restricting the change of PEB charging state, which can be easily proved, and the introduce of $u_{n,k}$ and $v_{n,k}$ makes the constraints linear; constraints (10) describe the charging demand constraints of PEBs. The recharged energy of PEBs during each charging between the minimum charging demands and the available battery capacity, as shown in (13).

$$\left(\Delta SOC_{n,i_n}^{\text{PEB}} + SOC_{\min}^{\text{PEB}} - SOC_{n,i_n}^{\text{PEB}}\right) S_n^{\text{PEB}} \leq$$
$$\sum_{k \in P(i_n)} c_{n,k}^{\text{PEB}} P_c^{\text{PEB}} \eta_c^{\text{PEB}} \Delta t \leq \left(1 - SOC_{n,i_n}^{\text{PEB}}\right) S_n^{\text{PEB}},$$
$$i_n \in I(n), n \in N \quad (13)$$

Based on (13), we utilize all the $SOC_{n,1}^{\text{PEB}}$ and $\Delta SOC_{n,i_n}^{\text{PEB}}$ to estimate the other $SOC_{n,i_n}^{\text{PEB}}$. Then, (13) can be rewritten in form of accumulation, i.e., (11). In (11), $i_n \leq j_n$ means parking $i_n$ occurs no later than parking $j_n$. And the middle part of each inequality in (11) represents the total recharged energy of PEB $n$ during $j_n$ times parking. The first and last parts of each inequality in (11) respectively represent the lower and upper limits of total recharged energy of PEB $n$ after $j_n$ times charging processes. Note that (13) and (11) both require $SOC_{\text{mix}}^{\text{PEB}} + \Delta SOC_{n,i_n}^{\text{PEB}} \leq 1, n \in N$.

The optimal charging state matrix of PEBs $\mathbf{C}^{\text{PEB}}$ can be obtained by solving *Model A*. And $\mathbf{C}^{\text{PEB}}$ should be converted into the corresponding charging state matrix of charging piles $\mathbf{C}^{\text{FCP}}$ before implementing the control strategy. This step can be achieved by checking the numbers of each PEB and the charging pile the PEB connected to.

*4.2.2 Optimisation Model for Coordinated PEB Charging Scenario with ESS:* Here, the optimisation model (*Model B*) is utilized when ESS is installed and charging loads of PEBs are controllable. To ensure the global optimality of the control strategy, PEB charging strategy and ESS charging and discharging strategy are optimised integratedly in the model. The objective function is to minimise the electricity purchase costs ($EPC$), the life expenditure costs of ESS ($ESSC$) and the equivalent capacity charge ($ECC$) in the following optimising time horizon. The whole model is formulated as (14)-(24).

$$\min EPC + ESSC + ECC \quad (14)$$

where $ECC$ is calculated by (3) and

$$EPC = \sum_{k \in K}\left(\sum_{n \in N} c_{n,k}^{\text{PEB}} P_c^{\text{PEB}}\right)\Delta t \pi_k^{\text{TOU}}$$
$$+ \left(\sum_{k \in K} P_{c,k}^{\text{ESS}} - \sum_{k \in K} P_{d,k}^{\text{ESS}}\right)\Delta t \pi_k^{\text{TOU}} \quad (15)$$

$$ESSC = \frac{\pi^{ESS}}{n^{ESS}} \sum_{k \in K} P_{c,k}^{\text{ESS}} \eta_c^{\text{ESS}} \Delta t \quad (16)$$

subject to:

$$(5)\text{-}(11) \quad (17)$$

$$P_{\text{peak}} \geq \sum_{n \in N} c_{n,k}^{\text{PEB}} P_c^{\text{PEB}} + P_{c,k}^{\text{ESS}} - P_{d,k}^{\text{ESS}} + L_k,$$
$$k \in K \quad (18)$$

$$0 \leq P_{c,k}^{\text{ESS}} \leq P_{c,\max}^{\text{ESS}}, k \in K \quad (19)$$

$$0 \leq P_{d,k}^{\text{ESS}} \leq P_{d,\max}^{\text{ESS}}, k \in K \quad (20)$$

$$SOC_k^{\text{ESS}} + P_{c,k}^{\text{ESS}} \eta_c^{\text{ESS}} \Delta t - P_{d,k}^{\text{ESS}} \Delta t / \eta_d^{\text{ESS}}$$
$$= SOC_{k+1}^{\text{ESS}}, k \in K \setminus \{\text{card}(K)\} \quad (21)$$

$$SOC_{\min}^{\text{ESS}} \leq SOC_k^{\text{ESS}} \leq 1, k \in K \quad (22)$$

$$P_{c,k}^{\text{ESS}} + P_{d,k}^{\text{ESS}} \leq \max\left(P_{c,k}^{\text{ESS}}, P_{d,k}^{\text{ESS}}\right), k \in K \quad (23)$$

$$\sum_{k \in K}\left(P_{c,k}^{\text{ESS}} \eta_c^{\text{ESS}} + P_{d,k}^{\text{ESS}} / \eta_d^{\text{ESS}}\right) = 0 \quad (24)$$

In the above model, equation (15) calculates the electricity purchase costs for PEBFCS with ESS. In (15), $P_{c,k}^{\text{EES}}$ and $P_{d,k}^{\text{EES}}$ denote the charging power and discharging power of ESS at time interval $k$, respectively. Equation (16) approximately calculates the ESS life expenditure costs during the optimising time horizon through the recharged energy, where, $\pi^{\text{ESS}}$, $n^{\text{ESS}}$ and $\eta_c^{\text{ESS}}$ are the unit price of ESS, the number of charge-discharge life cycle of ESS and the charging efficiency of ESS, respectively. The peak load $P_{\text{peak}}$ here is constrained by (18) and the optimal solution must make (25) hold (similar to constraint (4) in *Model A*).

$$P_{\text{peak}} = \max_{k \in K}\left(\sum_{n \in N} c_{n,k}^{\text{PEB}} P_c^{\text{PEB}} + P_{c,k}^{\text{ESS}} - P_{d,k}^{\text{ESS}} + L_k\right) \quad (25)$$

For constraints (19)-(24), (19) and (20) restrict the charging power and discharging power of ESS within the maximum powers, respectively; (21) describe the energy state transition of ESS; (22) are the SOC range constraints for ESS; (23) describe the mutual exclusion of charging and discharging states of ESS, i.e., avoid simultaneous charging and discharging of ESS; (24) is the energy balance constraint of ESS, since it is usually expected that the initial and final SOCs are same.

In above constraints, nonlinear constraints (23) can be deleted from *Model B* without any sacrifice of optimality due to the charging and discharging efficiency. The detailed proof is omitted here and interested reader may need to refer to the appendix in [19], which is similar.

Solving *Model B*, the optimal charging state matrix of PEBs $\mathbf{C}^{\text{PEB}}$, the optimal vector of charging power of ESS $\mathbf{P}_{\text{c}}^{\text{ESS}}$ and the optimal vector of discharging power of ESS $\mathbf{P}_{\text{d}}^{\text{ESS}}$ are accordingly obtained.

*4.2.3 Optimisation Model for Uncoordinated PEB Charging Scenario with ESS:* In this subsection's optimisation model (*Model C*), ESS is taken into account and charging loads of PEBs cannot be scheduled. The expression of the objective function is same as (14), but the charging profile of PEBs, i.e., $\sum_{n \in N} c_{n,k}^{\text{PEB}} P_{\text{c}}^{\text{PEB}}$, is input parameters here. The problem is modeled as (26) and (27).

$$\min EPC + ESSC + ECC \quad (26)$$

where $EPC$, $ESSC$ and $ECC$ are calculated by (15), (16) and (3), respectively.
subject to:
$$(18)\text{-}(22), (24). \quad (27)$$

The optimal vectors of charging and discharging powers of ESS, i.e., $\mathbf{P}_{\text{c}}^{\text{ESS}}$ and $\mathbf{P}_{\text{d}}^{\text{ESS}}$, can be obtained by solving *Model C*.

### 4.3. Heuristics-based Method for Coordinated PEB Charging Scenario with ESS

In the previous text, all the formulated models, i.e., *Model A*, *Model B*, and *Model C*, are mixed integer linear programming (MILP) models. However, different from *Model C*, *Model A* and *Model B* bear the extra computation burdens caused by the introduced auxiliary variables $u_{n,k}$ and $v_{n,k}$. To fix this, herein, we further propose a heuristics-based method.

Firstly, we solve *Model A* without regard to constraints (7) and (8) and obtain a charging state matrix of PEBs $\tilde{\mathbf{C}}^{\text{PEB}}$ (elements: $\tilde{c}_{n,k}^{\text{PEB}}$). According to $\tilde{\mathbf{C}}^{\text{PEB}}$ and the PEB departure time table, we can extract approximate charging time of all the PEBs during their each parking, denoted by $CT_{n,i_n}$, in the following optimising time horizon (see (28)). Also, we treat $\hat{\mathbf{C}}^{\text{PEB}} = \text{sum}(\tilde{\mathbf{C}}^{\text{PEB}})$ (elements: $\hat{c}_k^{\text{PEB}}$; sum($\mathbf{X}$) is a row vector whose elements are the sum of every column of $\mathbf{X}$) as the guideline of the number of PEB on charge at each time interval.

$$CT_{n,i_n} = \sum_{k=a_{i,i_n}+1}^{l_{n,i_n}-1} \tilde{c}_{n,k}^{\text{PEB}} \quad (28)$$

Secondly, inspired by [23] and [24], we define the laxity $LX_{n,k}$ (see (29)), which is used to describe the flexibility of the PEB charging. Note that 1) the laxity is defined for parked PEBs that wait for charging, otherwise (if the PEB is on charge or has been charged), the laxity is invalid; 2) for a new time interval, laxity values should be updated; 3) for a given $k$, a specific $i_n$ can be found if PEB $n$ parks at the PEBFCS, and $a_{n,i_n}$ and $l_{n,i_n}$ are therefore determined.

$$LX_{n,k} = l_{n,i_n} - CT_{n,i_n} - k, a_{n,i_n} < k < l_{n,i_n} \quad (29)$$

Based on $T_{n,i_n}$, $\hat{c}_k^{\text{PEB}}$ and $LX_{n,k}$, a heuristic algorithm is then developed to dispatch the PEB charging with the continuous charging constraints involved and generate the final charging state matrix of PEBs $\mathbf{C}^{\text{PEB}}$, given in Algorithm 1.

| **Algorithm 1. PEB Charging Dispatching** |
|---|
| 1:   **Initialization:** Set $\mathbf{C}^{\text{PEB}} = \mathbf{0}$. |
| 2:   **for** $k = 1$ **to** $K$ **do** |
| 3:     Find all the parked PEBs that are waiting for charging and compute their $LX_{n,k}$. |
| 4:     **if** $LX_{k,n} = 0$ **then** |
| 5:       Set $c_{n,k}^{\text{PEB}} = c_{n,k+1}^{\text{PEB}} = \cdots = c_{n,k+CT_{n,i_n}-1}^{\text{PEB}} = 1$. |
| 6:     **end if** |
| 7:     Sort PEBs according to the increasing $LX_{n,k} > 0$ and for PEBs with same $LX_{n,k}$, sort them according to the decreasing $CT_{n,i_n}$. |
| 8:     **if** $\text{sum}(\mathbf{C}^{\text{PEB}}) < \hat{\mathbf{C}}^{\text{PEB}}$ **then** |
| 9:       Select $\hat{\mathbf{C}}^{\text{PEB}} - \text{sum}(\mathbf{C}^{\text{PEB}})$ PEBs according to the order in line 7 and set $c_{n,k}^{\text{PEB}} = c_{n,k+1}^{\text{PEB}} = \cdots = c_{n,k+CT_{n,i_n}-1}^{\text{PEB}} = 1$. |
| 10:    **end if** |
| 11:  **end for** |

Armed with $\mathbf{C}^{\text{PEB}}$ got by Algorithm 1, we are then able to skip auxiliary variables $u_{n,k}$ and $v_{n,k}$ and seek for the ESS charging and discharging strategy through *Model C*. Note that, in the heuristics-based method, the control strategies of PEBs and ESS are generated separately and the PEB coordinated charging is dispatched heuristically so that the final strategy is not necessarily the optimal strategy.

## 5. Case Studies
### 5.1. Parameter Settings

The proposed strategies are tested on a PEBFCS with 10 fast charging piles, which provides charging service to a loop PEB line with 24 PEBs. The settings of the PEBFCS are on the basis of a practical PEBFCS in Chongqing, China. ESS in the PEBFCS is composed of lithium titanate batteries, and the discount rate, the life cycle, and the capacity charge for the station are respectively set as $\alpha = 5\%$, $\gamma = 50$ year and $\pi^{\text{Cap}} = 14847$ RMB/kW [25]. The circle length of the PEB route is 50 km. According to the central limit theorem, we assume the average speeds of PEBs (km/h) follow Gaussian distribution $\text{N}(50,5^2)$, and all $\Delta SOC_{n,i_n}^{\text{PEB}}$ (kWh) follow

**Table 1** Parameters settings of PEBs and ESS

| $S_n^{PEB}$ (kWh) | $\eta_c^{PEB}$ | $P_c^{PEB}$ (kW) | $SOC_{min}^{PEB}$ | $S^{ESS}$ (kWh) | $P_{c,max}^{ESS}$ (kW) | $P_{d,max}^{ESS}$ (kW) | $\eta_c^{ESS}$ | $\eta_d^{ESS}$ | $n^{ESS}$ | $SOC_{min}^{ESS}$ |
|---|---|---|---|---|---|---|---|---|---|---|
| 324 | 0.92 | 117 | 0.2 | 800 | 1000 | 1000 | 0.92 | 0.92 | 15000 [26] | 0.2 |

**Table 2** PEB departure time-table

| Time | Number* | Interval (min) | Time | Number* | Interval (min) |
|---|---|---|---|---|---|
| 06:00-07:00 | 12 | 5 | 15:00-16:00 | 6 | 10 |
| 07:00-08:00 | 12 | 5 | 16:00-17:00 | 12 | 5 |
| 08:00-09:00 | 12 | 5 | 17:00-18:00 | 12 | 5 |
| 09:00-10:00 | 12 | 5 | 18:00-19:00 | 4 | 15 |
| 10:00-11:00 | 12 | 5 | 19:00-20:00 | 4 | 15 |
| 11:00-12:00 | 12 | 5 | 20:00-21:00 | 4 | 15 |
| 12:00-13:00 | 12 | 5 | 21:00-22:00 | 2 | 30 |
| 13:00-14:00 | 12 | 5 | 22:00-23:00 | 1 | 60 |
| 14:00-15:00 | 6 | 10 | Other | 0 | - |

*\* The number of PEBs depart the PEBFCS at each time interval.*

**Table 3** TOU electricity prices

| Time | | Price (RMB/kWh) |
|---|---|---|
| Valley | 23:00-7:00 | 0.3818 |
| Shoulder | 7:00-10:00, 15:00-18:00, 21:00-23:00 | 0.8395 |
| High | 10:00-15:00, 18:00-21:00 | 1.3222 |
| Peak | 11:00-13:00, 20:00-21:00 | 1.4409 |

Gaussian distribution $N(70, 7^2)$. Specific parameters of PEBs and ESS are listed in Table 1, where the PEB parameters partially refer to those of BYD K9 pure electric bus [27]. The PEB departure time-table is listed in Table 2 and the TOU electricity tariffs are given in Table 3 [28]. The typical load profile in [29] is selected as the other load curve in the case. The optimising time horizon is set as 24h and card$(K)$ is 288, i.e., $\Delta t=5$ min.

Based on the parameters settings above, we carry out the simulations when the unit price of ESS, i.e., $\pi^{ESS}$, is respectively 8000 RMB/kWh, 6000 RMB/kWh, 4000 RMB/kWh, and 2000 RMB/kWh.

### 5.2. Results and Analysis

Numerical simulation results under uncoordinated and coordinated PEB charging scenarios are respectively summarized in Tables 4 and 5 (including the optimal and heuristics-based strategies), where AOC is the abbreviation of annual operation costs (including the ESS expenditure costs, equivalent capacity charge and electricity bills per year). Typical daily load profile, SOC curve of ESS and charging/discharging power curves of ESS under uncoordinated PEB charging scenario are given in Figs. 3 and 4, and the corresponding profiles under coordinated PEB charging scenario are shown in Figs. 5 and 6. All the problems were solved by the CPLEX package [30] on a laptop with an Intel Core i5 processor and 4 GB random-access memory.

From Tables 4 and 5, by comparing with the results without ESS, it can be seen that the proposed control strategies effectively improve AOC of PEBFCS and peak loads both under coordinated and uncoordinated PEB charging scenarios. And as the price of ESS falls, AOC decreases. According to whether charging loads of PEBs are controllable and whether ESS is considered, there are four different AOCs under each price of ESS. These four AOCs demonstrate that the combination of coordination of PEB and ESS achieves the best optimality. Besides, when loads of PEB and ESS are both coordinated, it is observed that the controlled peak load first decreases and then rebounds as the price of ESS declines from 8000 RMB/kWh to 2000 RMB/kWh (both using optimal and heuristics-based strategies). The reason is that electricity price arbitrage makes more profits than to decrease the capacity charge when the price of ESS is sufficiently low. While under the uncoordinated PEB charging scenario, the controlled peak load remains constant with the change of the ESS price because all the capacity of ESS is used to shave the peak PEB charging loads during the high and peak TOU price periods, which brings larger benefits than ESS costs. Regarding the different results of optimal and heuristics-based strategies in Table 5, it can be observed that the heuristics-based strategy gives rise to a slight increasing of AOC and peak loads, but has a distinct advantage in computation speed. In practice, if $\Delta t=5$ min, both optimal and heuristics-based strategies are applicable for a PEBFCS of such size, because all the computation time is much shorter than $\Delta t$. If the PEBFCS size increases or $\Delta t$ decreases, the computation time could matter and the heuristics-based strategy is probably a more appropriate choice. For example, we double the PEBFCS size, i.e., the PEB number increases to 48 and the departure numbers in Table 2 are all doubled, and let $\Delta t=3$ min, i.e., card$(K)$ is 480, then the computation time of optimal strategy exceeds

**Table 4** Numerical simulation results under the uncoordinated PEB charging scenario

| Price of ESS (RMB/kWh) | AOC (without ESS) (RMB) | AOC (with ESS) (RMB) | Reduction of AOC (%) | Peak Load (without ESS) (kW) | Peak Load (with ESS) (kW) | Reduction of Peak Load (%) | The Average Computation Time (s) |
|---|---|---|---|---|---|---|---|
| 8000 | 3800453 | 3474289 | 8.58  | 1366.4 | 1045.1 | 23.51 | 5.18 |
| 6000 |         | 3436731 | 9.57  |        | 1045.1 | 23.51 | 9.12 |
| 4000 |         | 3374498 | 11.21 |        | 1045.1 | 23.51 | 9.34 |
| 2000 |         | 3312229 | 12.85 |        | 1045.1 | 23.51 | 9.23 |

**Table 5** Numerical simulation results under the coordinated PEB charging scenario (optimal/heuristics-based)

| Price of ESS (RMB/kWh) | AOC (without ESS) (RMB) | AOC (with ESS) (RMB) | Reduction of AOC (%) | Peak Load (without ESS) (kW) | Peak Load (with ESS) (kW) | Reduction of Peak Load (%) | The Average Computation Time (s) |
|---|---|---|---|---|---|---|---|
| 8000 | 1401710 | 1314657/1324911 | 6.12/5.48  | 1305.3 | 1057.1/1057.1 | 19.01/19.01 | 19.02/0.19 |
| 6000 |         | 1284581/1293059 | 8.33/7.75  |        | 1009.2/1023.0 | 22.68/21.63 | 20.31/0.19 |
| 4000 |         | 1236876/1243737 | 11.76/11.24|        | 1009.2/1023.2 | 22.68/21.61 | 22.24/0.21 |
| 2000 |         | 1181943/1187108 | 15.68/15.31|        | 1015.6/1024.7 | 22.19/21.50 | 24.51/0.21 |

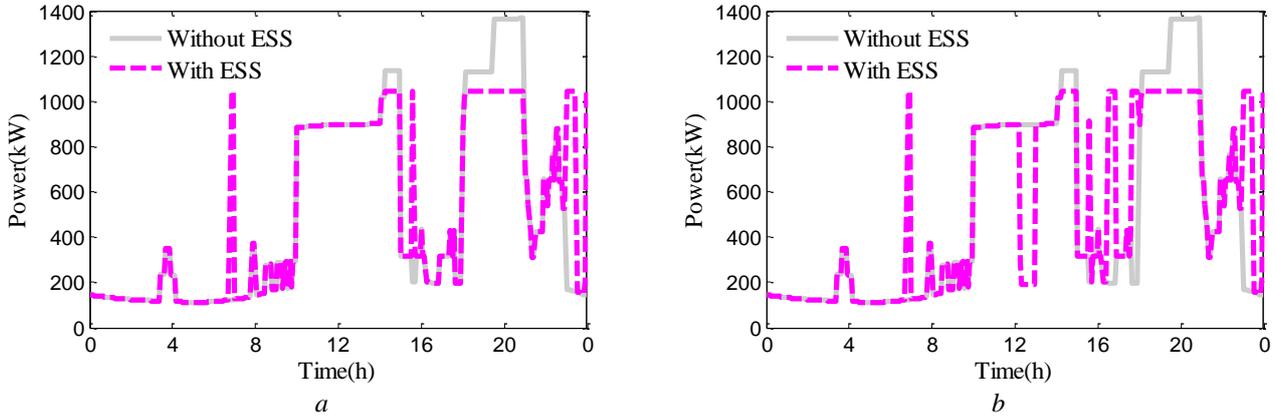

*Fig. 3.* *Typical daily load profiles of the local distribution transformer under uncoordinated PEB charging scenario*
*(a)* Price of ESS: 8000 RMB/kWh, *(b)* Price of ESS: 6000 RMB/kWh, 4000 RMB/kWh, and 2000 RMB/kWh

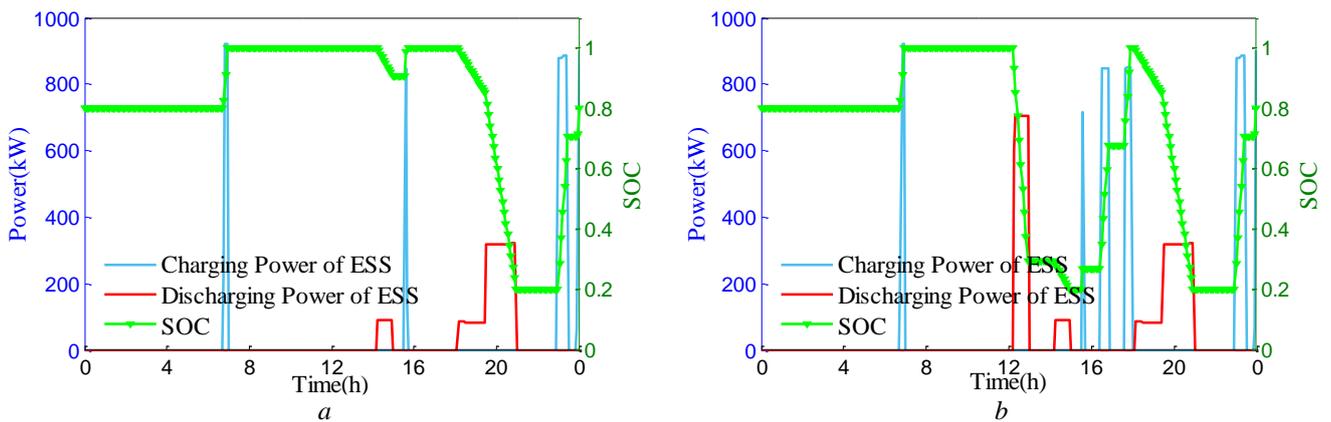

*Fig. 4.* *Typical daily SOC curves and charging/discharging power curves of ESS under uncoordinated PEB charging scenario*
*(a)* Price of ESS: 8000 RMB/kWh, *(b)* Price of ESS: 6000 RMB/kWh, 4000 RMB/kWh, and 2000 RMB/kWh

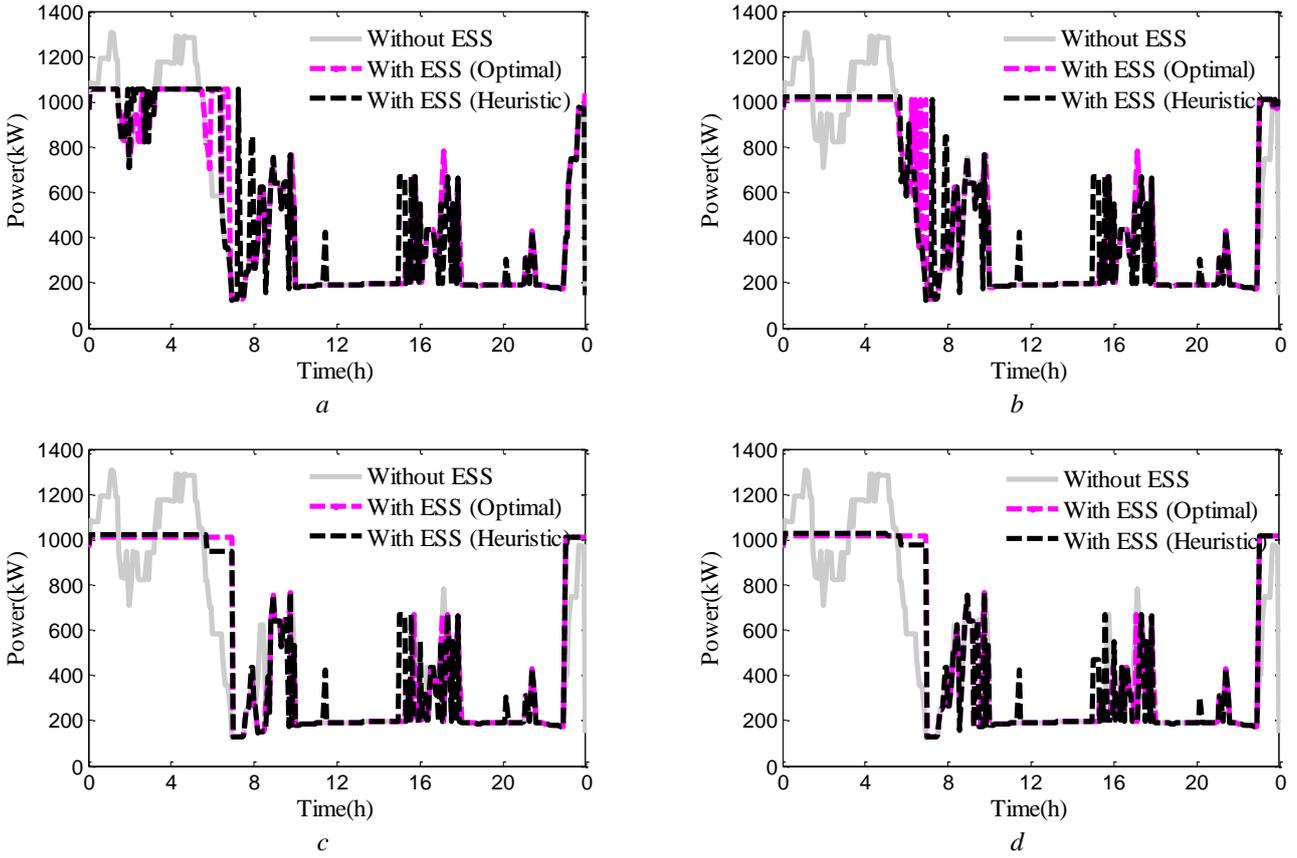

*Fig. 5.* *Typical daily load profiles of the local distribution transformer under coordinated PEB charging scenario*
*(a)* Price of ESS: 8000 RMB/kWh, *(b)* Price of ESS: 6000 RMB/kWh, *(c)* Price of ESS: 4000 RMB/kWh, *(d)* Price of ESS: 2000 RMB/kWh

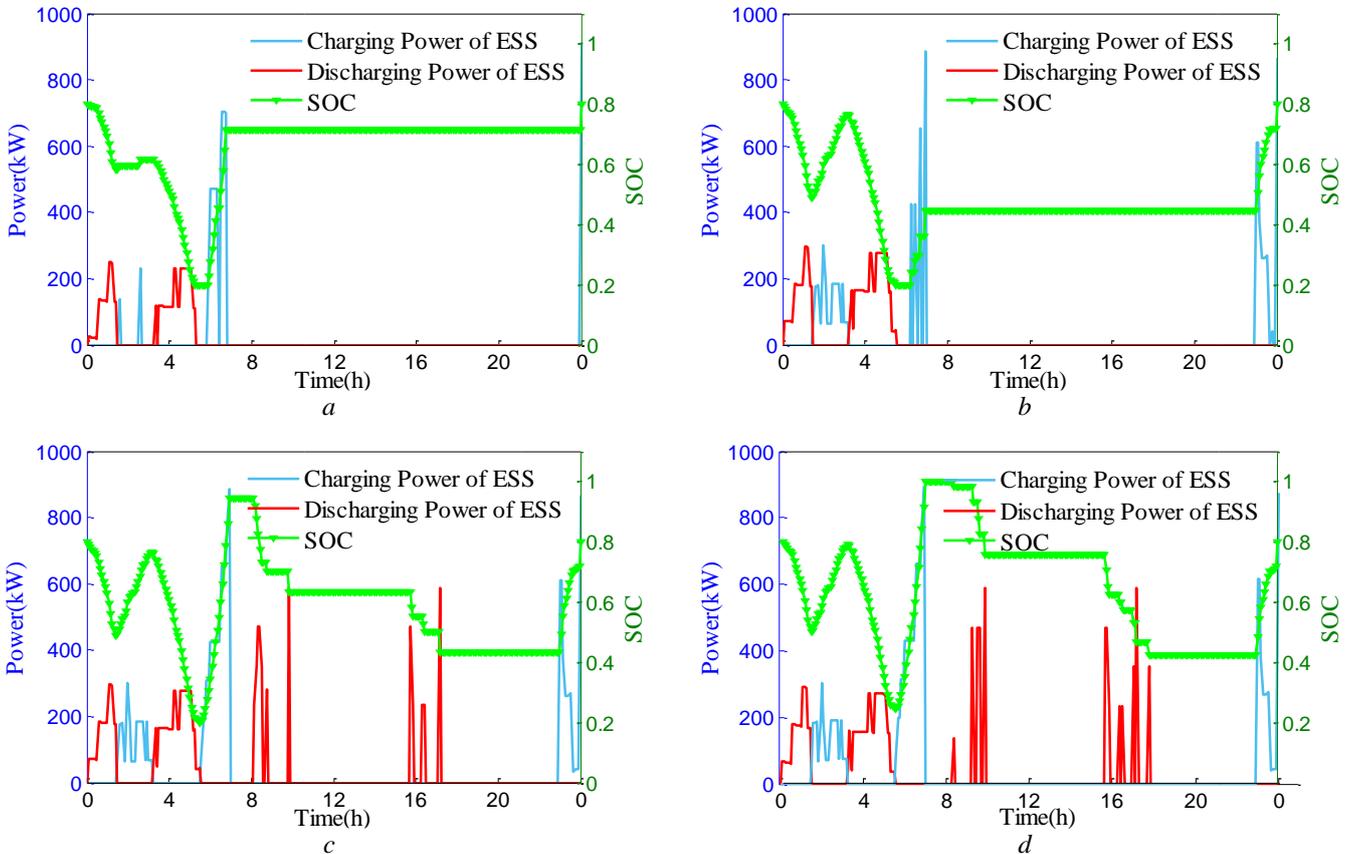

*Fig. 6.* *Typical daily SOC curves of ESS and charging/discharging power curves of ESS under coordinated PEB charging scenario (optimal)*

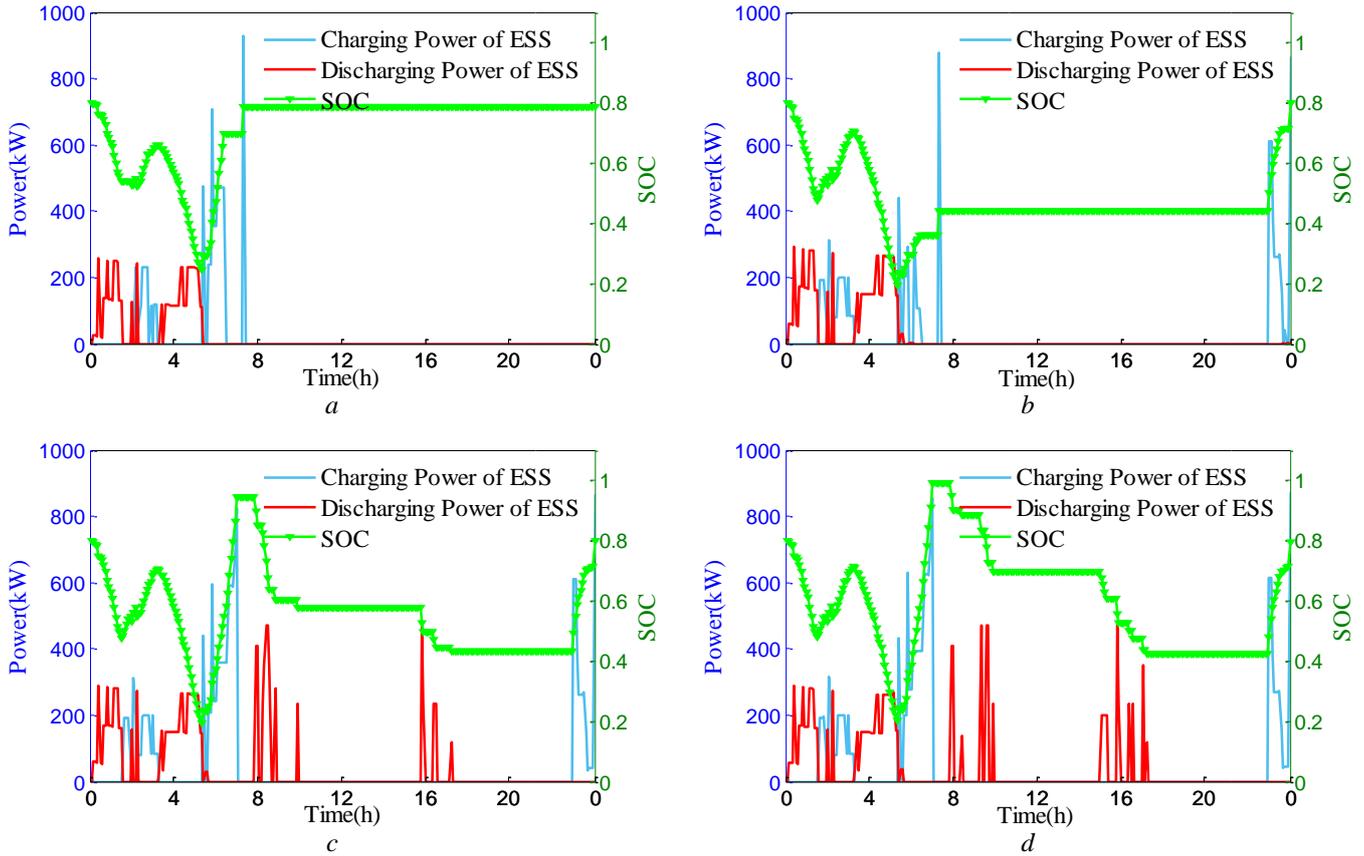

*(a)* Price of ESS: 8000 RMB/kWh, *(b)* Price of ESS: 6000 RMB/kWh, *(c)* Price of ESS: 4000 RMB/kWh, *(d)* Price of ESS: 2000 RMB/kWh

**Fig. 7.** *Typical daily SOC curves of ESS and charging/discharging power curves of ESS under coordinated PEB charging scenario (heuristics-based)*
*(a)* Price of ESS: 8000 RMB/kWh, *(b)* Price of ESS: 6000 RMB/kWh, *(c)* Price of ESS: 4000 RMB/kWh, *(d)* Price of ESS: 2000 RMB/kWh

$\Delta t=3$ min but the time of heuristics-based strategy is within 5s. Thus, under this scenario with larger station size and more frequent strategy update rate, heuristics-based strategy still works while the optimal strategy becomes unpractical due to over-long calculation time.

Comparing the load profiles in Figs. 3 and 5, we can see that 1) when there is no ESS, the coordinated PEB charging shifts the peak loads from day (high electricity prices) to night (low electricity prices); 2) when ESS is taken into account, the control strategies, including the optimal and heuristics-based (only for coordinated PEB charging scenario) strategies, smooth the load profiles under both two scenarios.

Recall that under uncoordinated PEB charging scenario, ESS is made full use of and the main restriction of further costs decreasing is the capacity. Due to this, daily load profile, SOC curve of ESS and charging/discharging power curves of ESS are identical when the price of ESS is respectively 6000 RMB/kWh, 4000 RMB/kWh and 2000 RMB/kWh (see Figs. 3 and 4). The reason, why the profiles under the ESS price 8000 RMB/kWh differ, is that charging ESS in the shoulder TOU price period and discharging ESS in the peak TOU price period are not profitable at such an ESS price (can be observed by comparing subfigures in Figs. 3 and 4). In other words, there is a threshold value of ESS price, which is between 6000 RMB/kWh and 8000 RMB/kWh, and PEBFCS can make profits through electricity price arbitrage between shoulder and peak TOU prices if the ESS price is lower than the value, otherwise larger electricity price difference, e.g., the price difference between peak and valley TOU prices, is needed.

Under coordinated PEB charging scenario, the subfigures in Fig. 5 illustrate that the total load profile of PEBFCS with ESS becomes more and more flatter as the price of ESS falls, and the subfigures in Figs. 6 and 7 show that the usage frequency of ESS trends to increase as the price of ESS falls. Besides, it is observed that charging and discharging of ESS both occur in the valley period of electricity price (see Figs. 6 and 7). And as a result, the night peak loads are further flattened, which implies that economic losses caused by energy consumption during the charging and discharging process are less than the reduction of capacity charge. Also, Figs. 5-7 illustrate that the load profiles, SOC curves of ESS, and charging/discharging power curves of ESS are similar when using optimal and heuristics-based strategies.

### 5.3. Sensitivity Analysis of ESS Capacity

We study the impact of the ESS capacity on AOC when using the optimal control strategy under PEB coordinated charging scenario (the impacts are similar for different strategies and scenarios). The ESS prices 6000 RMB/kWh and 4000 RMB/kWh are selected for the sensitivity analysis, and the results are given in Fig. 8. It can be seen that AOC decreases and converges to a constant as the ESS capacity increases. However, when the AOC stops declining, the extra capacity of idle ESS will lead to unnecessary investment, which is not included in EOA. So, in practice, the ESS capacity at the stop point of AOC decreasing is most appropriate for installation. In Fig. 8, the best ESS capacity at ESS prices 6000 RMB/kWh and 4000 RMB/kWh are both between 1000 kWh and 1200 kWh.

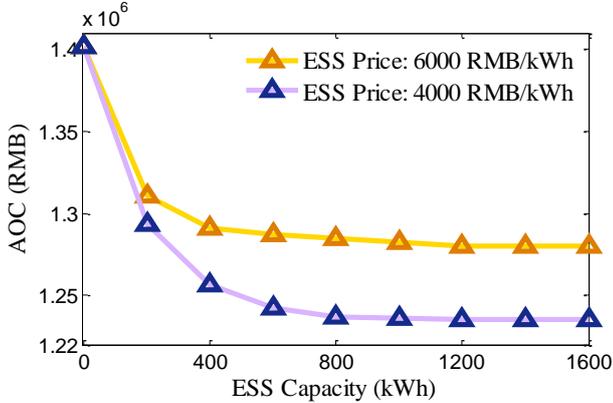

***Fig. 8.*** *Curves of sensitivity analysis of ESS capacity*

## 6. Conclusion

This paper proposes coordinated charging and discharging strategies for a PEBFCS with ESS to optimise the economic benefits. The mathematical models are respectively formulated when the PEB charging loads are controllable or not. And when PEB charging loads are controllable, i.e., under coordinated PEB charging scenario, a heuristics-based strategy is further proposed to enhance the computation efficiency with a little sacrifice of optimality. We validate the effectiveness of the proposed strategies under multiple ESS prices through case studies, and analyze the impacts of ESS capacity on the PEBFCS operation costs. Further research includes the cooperation strategy for several PEBFCSs with ESS.

## 7. Acknowledgments

This work was supported in part by the National Natural Science Foundation of China (51477082) and the National Key Research and Development Program (2016YFB0900103).